\documentclass[11pt]{article}
\usepackage{graphicx,latexsym}

\usepackage{amsmath,amssymb}
\usepackage{latexsym}
\usepackage{graphicx}
\usepackage{algorithm}
\usepackage{amsthm}
\usepackage{authblk}

\newtheorem{theorem}{Theorem}
\newtheorem{proposition}{Proposition}

\newtheorem{definition}{Definition}
\newcommand{\pf}{{\it Proof.~~}}
\newcommand{\QED}{\hfill $\Box$}

\begin{document}

\title{Tree Edit Distance with Variables. Measuring the Similarity between
Mathematical Formulas}

\author[1]{Tatsuya Akutsu\thanks{Corresponding author. e-mail: takutsu@kuicr.kyoto-u.ac.jp}}
\author[1]{Tomoya Mori}
\author[2,3,4]{Naotoshi Nakamura}
\author[2,3]{Satoshi Kozawa}
\author[2,3,5]{Yuhei Ueno}
\author[2,3,5]{Thomas N. Sato\thanks{Corresponding author. e-mail: island1005@gmail.com}}
\affil[1]{Bioinformatics Center, Institute for Chemical Research, Kyoto University}
\affil[2]{The Thomas N. Sato BioMEC-X Laboratories,
Advanced Telecommunications Research Institute International (ATR)}
\affil[3]{Karydo TherapeutiX, Inc.}
\affil[4]{Center for Mathematical Modeling and Data Science, Osaka University}
\affil[5]{V-iCliniX Laboratory, Nara Medical University}


\maketitle

\begin{abstract}
In this article, we propose
\emph{tree edit distance with variables},
which is an extension of the tree edit distance to handle trees
with variables and has a potential application to measuring
the similarity between mathematical formulas, especially, those appearing
in mathematical models of biological systems.
We analyze the computational complexities of several variants
of this new model.
In particular, we show that the problem is NP-complete for ordered trees.
We also show for unordered trees that
the problem of deciding whether or not the distance is
0 is graph isomorphism complete but can be solved in polynomial time
if the maximum outdegree of input trees is bounded by a constant.
This distance model is then extended for measuring the difference/similarity
between two systems of differential equations, for which
results of preliminary computational experiments using biological models
are provided.
\end{abstract}

\section{Introduction}
\label{sec:intro}

In this article, we consider the problem of computing edit distance
between trees with variables.
This problem is motivated from studies on comparison of mathematical
formulas/models \cite{vittadello20,zhong19}.
For example, consider two functions $f(x,y,z)$ and $g(x,y,z)$ defined by:
\begin{eqnarray*}
f(x,y,z) & = & (x+y)*z,\\
g(x,y,z) & = & (x+z)*y.
\end{eqnarray*}
These two functions are essentially the same:
the former one is identical to the latter one by
replacing $y$ and $z$ with $z$ and $y$, respectively.
In addition, consider a function $h(x,y,z)$ defined by:
\begin{eqnarray*}
h(x,y,z) & = & z*(x+y).
\end{eqnarray*}
This function is also essentially the same as $f$ and $g$
because multiplication satisfies the commutative law.

\begin{figure}[ht]
\begin{center}
\includegraphics[width=12cm]{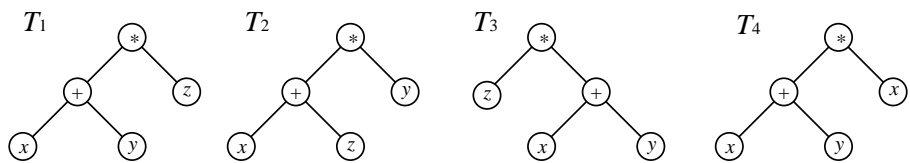}
\caption{Tree representations of mathematical expressions.}
\label{fig:tree-representation}
\end{center}
\end{figure}

In order to examine the identify/similarity of mathematical expressions,
\emph{tree edit distance} has been utilized
because mathematical expressions can often be represented as rooted trees,
where tree edit distance is a measure of dissimilarity between two rooted
trees \cite{akutsu10,bille05}.
For example, functions $f$, $g$, and $h$ can be respectively represented as
$T_1$, $T_2$, and $T_3$ shown in Fig.~\ref{fig:tree-representation},
If we ignore variable names assigned to leaves,
these trees are identical as unordered rooted trees.

However, considering variable names is important,
where variables are often referred to as species in biological models.
For example, consider a function $k$ defined by
\begin{eqnarray*}
k(x,y) & = & (x+y)*x.
\end{eqnarray*}
This function can be represented as a rooted tree $T_4$
in Fig.~\ref{fig:tree-representation}.
Although (unordered) tree structures of $T_1,\ldots,T_4$ are identical,
$k$ is clearly different from $f$, $g$, and $h$.
Therefore, variable names assigned to leaves should be taken into account.
In order to take variables names into account,
unification with commutative and/or associative laws has been studied
\cite{akutsu17,kapur92}.
However, unification is used to decide a kind of
identity
between two expressions
and thus does not give a similarity (or distance) measure.

Based on the above discussion, we introduce
\emph{tree edit distance with variables} in this article.
Before giving this new distance measure,
we briefly review the standard \emph{tree edit distance}.
Let $T_1$ and $T_2$ be two rooted trees in which each node
has a label from an alphabet $\Sigma$.
We consider two cases: both $T_1$ and $T_2$ are ordered trees,
and both $T_1$ and $T_2$ are unordered trees.
This distinction can be taken into account only when
we consider whether or not two trees are identical (i.e., isomorphic)
after tree editing operations.
The tree edit distance $d_0(T_1,T_2)$ between $T_1$ and $T_2$ is defined as
the cost of the minimum cost sequence of edit operations that
transforms $T_1$ to $T_2$,
where an operation is one of deletion of a node, 
insertion of a node, and change of the label of a node.
Then, we define the tree edit distance between two trees
with variables $T_1$ and $T_2$ 
by $dist(T_1,T_2)=\min_{\theta} dist_0(T_1 \theta,T_2 \theta)$,
where $\theta$ is a substitution
(i.e., a set of assignments of constants to variables).
See Section~\ref{sec:pre} for the precise definitions.

In this article, we analyze the computational complexities of
several variants/subcases of the tree edit distance problem with variables.
When discussing the complexity classes, we consider a decision
version of the problem: whether or not $dist(T_1,T_2) \leq d$ for
given $T_1$, $T_2$, and a given non-negative real number $d$.
The results are summarized in Table~\ref{tab:summary},
where `iso' asks whether $d(T_1,T_2)=0$,
`BD' means that the maximum \emph{outdegree} (i.e., the maximum number
of children) of both $T_1$ and $T_2$ is bounded by a constant.
P, NPC, and GIC mean that the target problem is 
polynomial-time solvable, NP-complete, and Graph Isomorphism complete
(i.e., as hard as the graph isomorphism problem under 
polynomial-time reduction), respectively.
It is interesting to see that the complexity substantially changes according
to introduction of variables.

\begin{table}[ht]
\caption{Summary of Theoretical Results}
\label{tab:summary}
\begin{center}
\begin{tabular}{|c|c|cccc|}
\hline
& $d_0(T_1,T_2)$ & iso & iso-BD & $d(T_1,T_2)$ & $d(T_1,T_2)$-BD\\
\hline
ordered   & P \cite{tai79} & P & P & NPC & NPC \\
& & (Prop.~\ref{prop:ordered-poly}) &
(Prop.~\ref{prop:ordered-poly}) &
(Thm.~\ref{thm:ordered-hard}) &
(Thm.~\ref{thm:ordered-hard}) \\
\hline
unordered & NPC \cite{zhang92} & GIC & P & NPC \cite{zhang92} & NPC \cite{zhang92} \\
& & (Thm.~\ref{thm:unordered-iso}) &
(Thm.~\ref{thm:unordered-iso}) &
$O(m_1^{m_2} \cdot 1.26^{n_1+n_2})$ time & \\
& & & & (Prop.~\ref{prop:unordered-exp}) & \\
\hline
\end{tabular}
\end{center}
\end{table}

We also extend the tree edit distance with variables
for computing the distance between two systems of first-order differential
equations.
Then, we develop practical methods for computing this new distance and
its variant using integer linear programming (ILP).
Furthermore, we conduct preliminary computational experiments on
these methods using several mathematical models of biological systems
obtained from the BioModels data repository (https://www.ebi.ac.uk/biomodels/).

\section{Preliminaries}
\label{sec:pre}

In this section, we review the precise definition of the tree edit distance
and then formally define the tree edit distance with variables. 

\begin{figure}[ht]
\begin{center}
\includegraphics[width=12cm]{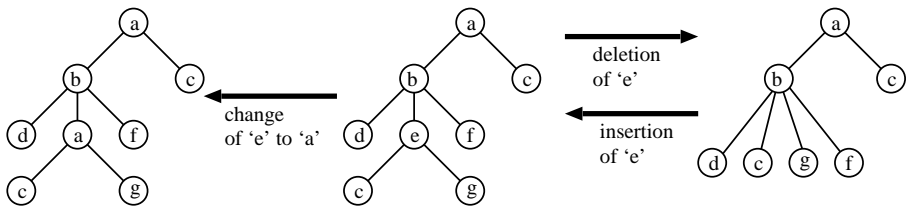}
\caption{Tree edit operations.}
\label{fig:edit-operations}
\end{center}
\end{figure}

Let $T_1$ and $T_2$ be two rooted trees in which each node
has a label from an alphabet $\Sigma$.
As mentioned in Section~\ref{sec:intro},
we consider two cases: both $T_1$ and $T_2$ are ordered trees,
and both $T_1$ and $T_2$ are unordered trees,
and this distinction can be taken into account only when
we consider whether or not two trees are identical
after tree edit operations.
We consider three kinds of \emph{edit operations} (see also Fig.~\ref{fig:edit-operations}):
\begin{description}
\item [Deletion:] Delete a non-root node $v$ in $T$ with parent $u$,
making the children of $v$ become children of $u$.
The children are inserted in the place of $v$ into the
set of the children of $u$.
\item [Insertion:] Inverse of delete.
Insert a node $v$ as a child of $u$ in $T$, making $v$ the parent
of some of the children of $u$.
\item [ChangeLabel:] Change the label of a node $v$ in $T$.
\end{description}
We assign a {\em cost} for each editing operation:
$\gamma(a,b)$ denotes the cost of changing a node with
label $a$ to label $b$,
$\gamma(a,\epsilon)$ denotes the cost of deleting a node labeled
with $a$, 
$\gamma(\epsilon,a)$ denotes the cost of inserting a node labeled
with $a$.
We assume that $\gamma(x,y)$ satisfies the conditions of distance
metric: $\gamma(x,x)=0$, $\gamma(x,y)=\gamma(y,x)$,
$\gamma(x,y) \geq 0$, and $\gamma(x,z) \leq \gamma(x,y) + \gamma(y,z)$.
Then, the \emph{edit distance} between $T_1$ and $T_2$ is defined as
the cost of the minimum cost sequence of edit operations 
that transforms $T_1$ to $T_2$ (precisely,
transforms $T_1$ to a tree identical to $T_2$).
It is well-known that this distance satisfies the conditions
of distance measure, in both ordered and unordered cases.

Here we define tree edit distance with variables.
Let $\Sigma$ be a set of constant symbols, 
where each constant is denoted by a lower-case letter
(e.g., $a,b,c,x,y,z,a_1,a_2$).
Let $\Lambda$ be a set of variables,
where each variable is denoted by an upper-case letter
(e.g., $X,Y,Z,X_1,X_2$).
A substitution is a set of variable-constant pairs,
$\theta = \{(X_1,x_1),(X_2,x_2),\ldots,(X_k,x_k)\}$,
where $X_i \neq X_j$ holds for all $i \neq j$ but
$x_i = x_j$ can hold for some $(i.j)$.
For a rooted tree $T$ and a substitution $\theta$,
$T \theta$ denotes the tree obtained by
changing variables appeared in $T$ to
constants according to $\theta$
(each $X_i$ is replaced with $x_i$).
Let $dist_0(T_1,T_2)$ be the standard tree edit distance
between $T_1$ and $T_2$ (i.e., distance between trees without variables).
We reasonably assume the following:
\begin{itemize}
\item Variable symbols appear only in leaves.
\item The sets of variables appearing $T_1$ and $T_2$ are disjoint.
\item Distinct variables in the same tree must be substituted to
distinct constants by $\theta$.
\item Every variable appearing in $T_1$ (resp., $T_2$)
is substituted to a constant symbol not appearing in $T_1$ or $T_2$
(because otherwise the cost of substituting a variable to a constant
would be 0, which is not appropriate for measuring the distance between
two mathematical expressions).
\end{itemize}
Then, we define the tree edit distance with variables as follows.

\begin{definition}
The tree edit distance with variables between $T_1$ and $T_2$
is
$$dist(T_1,T_2) = \min_{\theta} dist_0(T_1 \theta,T_2 \theta).$$
\end{definition}

\begin{figure}[ht]
\begin{center}
\includegraphics[width=10cm]{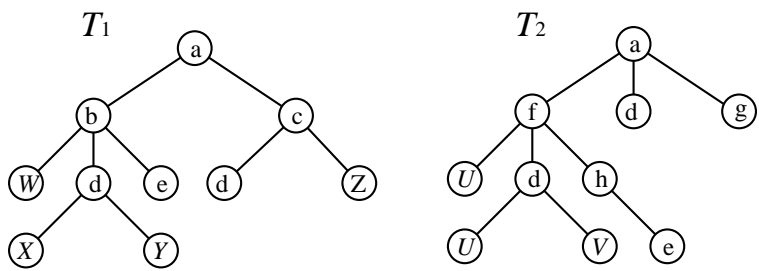}
\caption{Example of a tree pair. In this case, $dist(T_1,T_2)=5$ under
the unit cost model.}
\label{fig:example}
\end{center}
\end{figure}

For example, consider trees $T_1$ and $T_2$ shown in Fig.~\ref{fig:example}
and the unit cost model (i.e., $\gamma(x,y)=1$ for any $x \neq y$).
Then, $dist(T_1,T_2)=5$ (in both ordered and unordered cases) by
$\theta=\{(X,{\rm x}),(Y,{\rm y}),$
$(Z,{\rm z}),(W,{\rm w}),(U,{\rm x}),(V,{\rm y})\}$
and the following sequence of editing operations:
change the label of node `w' to `x',
insert node `h',
change the label of node `b' to `f',
delete node `c', and
change the label of node `z' to `g',
where we identify nodes by their labels.

As the basic property, the following is straightforward.

\begin{proposition}
For both ordered and unordered cases,
tree edit distance with variables satisfies the conditions of
distance measure.
\end{proposition}

\section{Ordered Trees}
\label{sec:ordered}

In this section, all trees are ordered trees,
which means that the children of each node are ordered 
from left to right and that this ordering must be preserved
among isomorphic trees.
For each tree $T$, $V(T)$ and $E(T)$ denote the sets of nodes and edges,
respectively.
We let $n_1 = |V(T_1)|$ and $n_2=|V(T_2)|$.
For each node (resp., vertex) $v$ in a tree (resp., in a graph),
$\ell(v)$ denotes the label of $v$.\footnote{We mainly use `nodes' for trees and
`vertices' for graphs.}

\begin{proposition}
For ordered trees, whether or not $dist(T_1,T_2)=0$
can be determined in polynomial time.
\label{prop:ordered-poly}
\end{proposition}
\pf
We construct an Euler string $str(T_i)$ \cite{akutsu06}
from each of a given tree $T_i$ using depth first search.
In constructing $str(T_i)$, we assign a unique integer number
from $1,2,\cdots$ as the label of a variable node every when we first encounter
the variable.
Then, it is straightforward to see $str(T_1)=str(T_2)$ if and only if
$dist(T_1,T_2)=0$.
\QED

\begin{figure}[ht]
\begin{center}
\includegraphics[width=12cm]{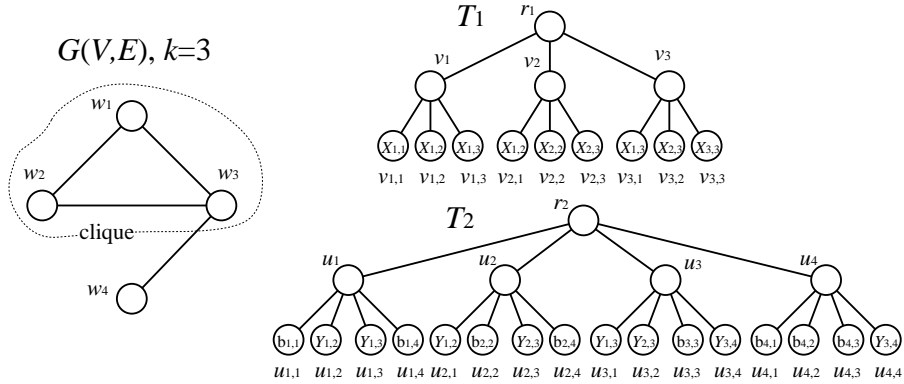}
\caption{Reduction from maximum clique to ordered tree edit distance with
variables, where only relevant labels are shown.}
\label{fig:ordered-hard}
\end{center}
\end{figure}

\begin{theorem}
For ordered trees, the tree edit distance problem with variables is NP-complete.
\label{thm:ordered-hard}
\end{theorem}
\pf
It is clear that the problem is in NP.
Then, we show a polynomial-time reduction from the maximum clique problem
(see also Fig.~\ref{fig:ordered-hard}).
The maximum clique problem is, given an undirected graph $G(V,E)$ and 
an integer $k$,
to decide whether or not there exists a complete subgraph (clique) of 
size (\#vertices) $k$ in $G(V,E)$,
where all vertices have the same label.
It is well-known that the problem is NP-complete.

From a given $k$, we construct $T_1$ as follows:
\begin{eqnarray*}
V(T_1) & =  & \{r_1\} \cup \{v_1, \ldots, v_k\} \cup
\left(\bigcup_{i \in \{1,\ldots,k\}} \{v_{i,1},\ldots,v_{i,k}\}\right),\\
E(T_1) & = & \left(\bigcup_{i \in \{1,\ldots,k\}} \{(r_1,v_i)\}\right)
~\cup~
\left(\bigcup_{i \in \{1,\ldots,k\}}
\{(v_i,v_{i,1}),\ldots,(v_i,v_{i,k})\}\right),\\
\ell(r_1) & = & \ell(v_1) = \cdots = \ell(v_k)={\rm a},\\
\ell(v_{i,j}) & = & \ell(v_{j,i})= X_{i,j} \mbox{ for all $i < j$},\\
\ell(v_{i,i}) & = & X_{i,i} \mbox{ for all $i$},
\end{eqnarray*}
where $X_{i,j} \neq X_{i',j'}$ for any $i \neq i'$ or $j \neq j'$.

From a given $G(V,E)$ with $V=\{w_1,\ldots,w_n\}$,
we construct $T_2$ as follows:
\begin{eqnarray*}
V(T_2) & =  & \{r_2\} \cup \{u_1, \ldots, u_n\} \cup
\left(\bigcup_{i \in \{1,\ldots,n\}} \{u_{i,1},\ldots,u_{i,n}\}\right),\\
E(T_2) & = & \left(\bigcup_{i \in \{1,\ldots,n\}} \{(r_2,u_i)\}\right)
~\cup~
\left(\bigcup_{i \in \{1,\ldots,n\}}
\{(u_i,u_{i,1}),\ldots,(u_i,u_{i,n})\}\right),\\
\ell(r_2) & = & \ell(u_1) = \cdots = \ell(u_n)={\rm a},\\
\ell(u_{i,j}) & = & \ell(u_{j,i})= Y_{i,j} \mbox{ for all $\{w_i,w_j\} \in E$ with $i < j$},\\
\ell(u_{i,j}) & = & {\rm b}_{i,j} \mbox{ for other nodes}.
\end{eqnarray*}
where $Y_{i,j} \neq Y_{i',j'}$ holds for any $i \neq i'$ or $j \neq j'$,
and all ${\rm b}_{i,j}$s are distinct constants.

Here, we note that $n_1=1+k+k^2$ and $n_2=1+n+n^2$.
Then, it is straightforward to see that
$G(V,E)$ has a clique of size $k$ if and only if
$dist(T_1,T_2) \leq n_2-n_1$ (under the unit cost model).

\begin{figure}[ht]
\begin{center}
\includegraphics[width=8cm]{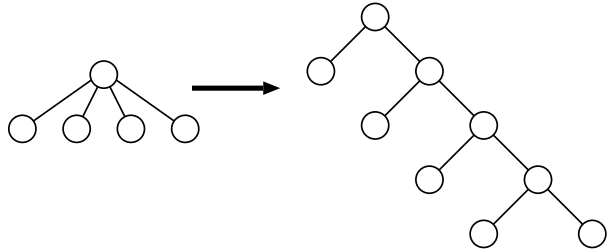}
\caption{Encoding of non-leaf node.}
\label{fig:star-encode}
\end{center}
\end{figure}

For the bounded case, it is enough to encode each non-leaf node as in Fig.~\ref{fig:star-encode}, where the details are omitted.
\QED

\section{Unordered Trees}
\label{sec:unordered}

In this section, all trees are unordered rooted trees.
The graph isomorphism problem is, given two undirected graphs
$G_1(V_1,E_1)$ and $G_2(V_2,E_2)$,
to decide whether or not there exists a bijection $\phi$ from $V_1$ to $V_2$
such that $\{u,v\} \in E_1$ if and only if $\{\phi(u),\phi(v)\} \in E_2$.
It is unclear that graph isomorphism is in P or NP-complete
(many researchers believe that it lies between P and NP-complete).
However, it is known that graph isomorphism can be solved in
polynomial time if the maximum degree of input graphs is bounded by
a constant \cite{grohe18}.

\begin{theorem}
For unordered trees, the problem of deciding
$dist(T_1,T_2)=0$ is graph isomorphism complete.
Furthermore, the problem can be solved in polynomial time
if the maximum outdegree of $T_1$ and $T_2$ is bounded by a constant.
\label{thm:unordered-iso}
\end{theorem}
\pf
First, we show that graph isomorphism is reduced to
the problem in polynomial time.
For each of $G_1$ and $G_2$,
we construct trees as for $T_2$ in the proof of Theorem~\ref{thm:ordered-hard}.
Then, it is straightforward to see that
$G_1$ and $G_2$ are isomorphic if and only if $dist(T_1,T_2)=0$.

\begin{figure}[ht]
\begin{center}
\includegraphics[width=12cm]{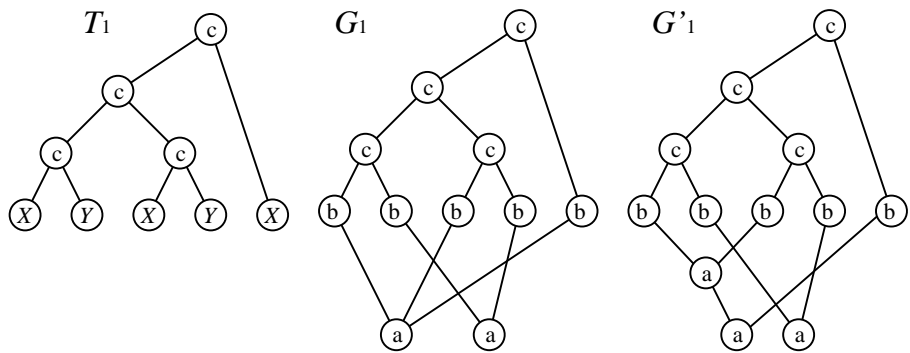}
\caption{Transformation from tree $T_1$ to graph $G_1$ of unbounded degree
and graph $G_1'$ of bounded degree.}
\label{fig:tree-to-graph}
\end{center}
\end{figure}

Next, we show that the problem is reduced to
graph isomorphism in polynomial time (see also Fig.~\ref{fig:tree-to-graph}).
Here, we consider without loss of generality (w.l.o.g.)
graph isomorphism over labeled graphs (because it is known that
labeled cases can be reduced to unlabeled cases
in polynomial time).
We show how to construct $G_1(V_1,E_2)$ from $T_1$,
where an identical construction can be used for $T_2$.
We construct $G_1(V_1,E_1)$ by adding vertices and edges to $T_1$ 
as follows.
For each variable $X$, we create a new vertex $v_X$ with constant label `a',
connect $v_X$ to all leaves in $T_1$ having label $X$,
and change the labels of these leaves to `b',
where `a' and `b' are constant symbols not appearing $T_1$ or $T_2$
(we use the same `a' and `b' for all variables in $T_1$ and $T_2$).
Then, it is straightforward to see that
$G_1$ and $G_2$ are isomorphic if and only if
$dist(T_1,T_2)=0$.

Finally, we show the last claim.
We modify the reduction shown above
(see also $G'_1$ in Fig.~\ref{fig:tree-to-graph}).
In order to reduce the degree of each of new vertices,
we make a copy of $T_1$ for each variable $X$.
Then, we keep leaves labeled with $X$ and
the nodes in the copy each of which has multiple children
whose descendants contain leaves labeled with $X$
(some child can be such a leaf),
and delete all other nodes in the copy.
Then, the maximum degree of the resulting graphs is bounded
by the maximum degree (not outdegree) of the input trees.
It is not difficult to see that
$G_1'$ and $G_2'$ are isomorphic if and only if
$dist(T_1,T_2)=0$.
\QED

Let $m_1$ and $m_2$ be the number of distinct variables appearing
in $T_1$ and $T_2$, respectively, where we assume w.l.o.g. $m_1 \leq m_2$.
Then, we have the following.

\begin{proposition}
$dist(T_1,T_2)$ can be computed in $O(m_2^{m_1} \cdot 1.26^{n_1+n_2})$ time.
\label{prop:unordered-exp}
\end{proposition}
\pf
Recall $dist(T_1,T_2) = \min_{\theta} dist_0(T_1 \theta,T_2 \theta)$.
Therefore, the problem can be solved by computing
$dist_0(T_1 \theta,T_2 \theta)$ for all essentially different $\theta$,
where ``essentially different'' $\theta_1$ and $\theta_2$
mean that $\theta_1$ and $\theta_2$ give distinct correspondences between
variables in $T_1$ and those in $T_2$.
The number of essentially different $\theta$ is clearly bounded by $m_2^{m_1}$.
Since the tree edit distance between two unordered trees
can be computed in $O(1.26^{n_1+n_2})$ time \cite{akutsu14},
the proposition holds.
\QED

\bigskip

\begin{figure}[ht]
\begin{center}
\includegraphics[width=8cm]{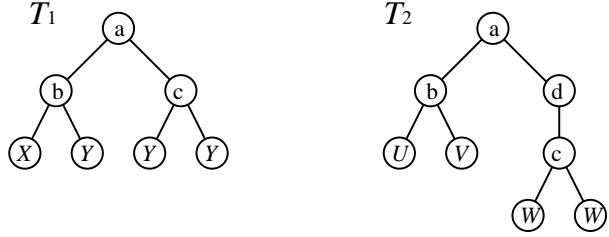}
\caption{Example for Proposition~\ref{prop:unordered-exp}.}
\label{fig:example-prop-exp}
\end{center}
\end{figure}

For example, consider $T_1$ and $T_2$ in Fig.~\ref{fig:example-prop-exp}.
In this case,
the following
are essentially different substitutions:
\begin{eqnarray*}
\theta_1 & = & \{(X,u),(Y,v),(U,u),(V,v),(W,w)\},\\
\theta_2 & = & \{(X,v),(Y,u),(U,u),(V,v),(W,w)\},\\
\theta_3 & = & \{(X,u),(Y,w),(U,u),(V,v),(W,w)\},\\
\theta_4 & = & \{(X,w),(Y,u),(U,u),(V,v),(W,w)\},\\
\theta_5 & = & \{(X,v),(Y,w),(U,u),(V,v),(W,w)\},\\
\theta_6 & = & \{(X,w),(Y,v),(U,u),(V,v),(W,w)\},
\end{eqnarray*}
where the number of substitutions is less than $m_2^{m_1}=3^2=9$.
Then, the minimum is attained for $\theta_3$ or $\theta_5$,
and we have
$$
dist(T_1,T_2) = dist_0(T_1 \theta_3,T_2 \theta_3) =
dist_0(T_1 \theta_5,T_2 \theta_5) = 2
$$
under the unit cost model
with the following sequence of editing operations:
inserting node `d' and then changing the label of the leftmost node `$Y$'.

\section{Comparison of Systems of Differential Equations}
\label{sec:equation}

In this section, we consider the problem of measuring the distance
between systems of differential equations.
We assume here that two systems of differential equations are given by
\begin{eqnarray*}
{\frac {dX_1}{dt}} & = & f_1(X_1,\ldots,X_{m_1}),\\
{\frac {dX_2}{dt}} & = & f_2(X_1,\ldots,X_{m_1}),\\
& \cdots & \\
{\frac {dX_{m_1}}{dt}} & = & f_{m_1}(X_1,\ldots,X_{m_1}),
\end{eqnarray*}
and
\begin{eqnarray*}
{\frac {dY_1}{dt}} & = & g_1(Y_1,\ldots,Y_{m_2}),\\
{\frac {dY_2}{dt}} & = & g_2(Y_1,\ldots,Y_{m_2}),\\
& \cdots & \\
{\frac {dY_{m_2}}{dt}} & = & g_{m_2}(Y_1,\ldots,Y_{m_2}),
\end{eqnarray*}
where some variables in some functions can be irrelevant.
In this article,
these systems are referred to as \emph{elementary systems},
and denoted by $S_X$ and $S_Y$, respectively.
As in the previous section, 
we assume that each function is represented as a tree,
$T^1_i$ for $f_i$ and $T^2_j$ for $g_j$,
where each tree is regarded as an unordered tree.
For simplicity,
we assume that all variables appearing in the right hand side
appear in the left hand side, although this restriction can be removed.
We assume w.l.o.g. that $m_1 \leq m_2$.
Let $n_1=\sum_{i=1}^{m_1} |V(T^1_i)|$ and
$n_2=\sum_{i=1}^{m_2} |V(T^2_i)|$.

As in the algorithm given in the proof of Proposition~\ref{prop:unordered-exp},
we consider all essentially different substitutions $\theta$
over variables $X_1,\ldots,X_{m_1}$ and $Y_1,\ldots,Y_{m_2}$.

Let $\hat{\theta}$ be the one-to-one mapping from $\{1,\ldots,m_1\}$
to $\{1,\ldots,m_2\}$ obtained from $\theta$ by the following rule:
\begin{itemize}
\item $X_i$ and $Y_j$ are substituted by the same constant by $\theta$ 
if and only if
$i$ is mapped to $j$ by $\hat{\theta}$ (i.e., $\hat{\theta}(i) = j$).
\end{itemize}
Let $I = \{j| \mbox{$\hat{\theta}(i)=j$ for some $i$}\}$ and
$O = \{1,\ldots,m_2\} \setminus I$.
Note that we can assume that $\hat{\theta}(i)$ is defined for
all $i=1,\ldots,m_1$ (otherwise, the distance would be larger).
Let $dist_0(\epsilon,T)=dist_0(T,\epsilon)$ be the cost
of deleting all nodes in $T$ (i.e., deleting the whole $T$).

\begin{definition}
The edit distance between two elementary systems $S_X$ and $S_Y$ is
$$Dist(S_X,S_Y) =
\min_{\theta} \left[
\sum_{i=1}^{m_1} dist_0(T^1_i \theta,T^2_{\hat{\theta}(i)} \theta) ~+~
\sum_{j \in O} dist_0(\epsilon,T^2_j) \right].$$
\label{def:elementary-dist}
\end{definition}

\begin{proposition}
$Dist(S_X,S_Y)$ can be computed in $O(m_2^{m_1} \cdot 1.26^{n_1+n_2})$ time.
\label{prop:elementary-dist}
\end{proposition}
\pf
We compute $Dist(S_X,S_Y)$ as in the algorithm given in the proof
of Proposition~\ref{prop:unordered-exp}.
We examine all essentially different $\theta$ and
compute 
$\sum_{i=1}^{m_1} dist_0(T^1_i \theta,T^2_{\hat{\theta}(i)} \theta)$ for
each $\theta$.
Since the computation time needed to compute this sum is bounded by
$$
\sum_{i=1}^{m_1} O(1.26^{|V(T^1_i)|+|V(T^2_{\hat{\theta}(i)})|})
~\leq~ O(1.26^{n_1+n_2}),
$$
the proposition holds.
\QED

\bigskip

\begin{figure}[ht]
\begin{center}
\includegraphics[width=12cm]{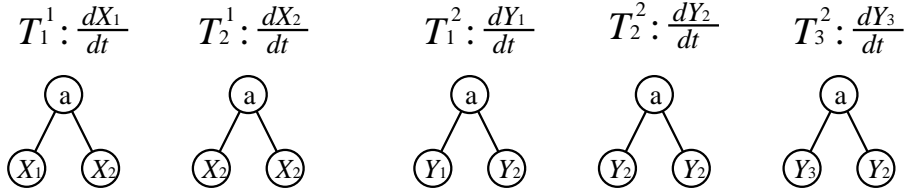}
\caption{Example for Proposition~\ref{prop:elementary-dist}.}
\label{fig:example-elementary}
\end{center}
\end{figure}

For example, consider trees given in Fig.~\ref{fig:example-elementary}.
Then, the minimum is attained for
$\theta = \{(X_1,y_3),(X_2,y_2),(Y_1,y_1),(Y_2,y_2),(Y_3,y_3)\}$
where $\hat{\theta}(1)=3$ and $\hat{\theta}(2)=2$,
and we have
$$
Dist(S_X,S_Y) = dist_0(T^1_1 \theta,T^2_3 \theta) +
dist_0(T^1_2 \theta,T^2_2 \theta) +
dist_0(\epsilon,T^2_1 \theta)
~=~ 
dist_0(\epsilon,T^2_1 \theta) ~=~ 3.
$$

The factor of $m_2^{m_1}$ in the above proposition may be too large
in practice.
Therefore, it is worthy to give another definition of the distance.
Let $\pi$ be a one-to-one mapping from $\{1,\ldots,m_1\}$
to $\{1,\ldots,m_2\}$, and let $O_{\pi}=
\{1,\ldots,m_2\} \setminus \{j| \mbox{$\pi(i)=j$ for some $i$}\}$.
Then, we define the \emph{pseudo edit distance} between two elementary systems
as below.

\begin{definition}
The pseudo edit distance between two elementary systems $S_X$ and $S_Y$ is
$$Pdist(S_X,S_Y) =
\min_{\pi} \left[
\sum_{i=1}^{m_1} dist(T^1_i,T^2_{\pi(i)}) ~+~
\sum_{j \in O_{\pi}} dist_0(\epsilon,T^2_j) \right].$$
\label{def:pseudo-dist}.
\end{definition}

\begin{proposition}
$Pdist(S_X,S_Y) \leq Dist(S_X,S_Y)$ holds for any $(S_X,S_Y)$, and
$Pdist(S_X,S_Y)$ can be computed in $O(poly(m_1,m_2) 1.26^{m_1+m_2})$ time.
\label{prop:pseudo-dist}
\end{proposition}
\pf
First note that, different from Definition~\ref{def:elementary-dist},
substitutions for $T^1_i$ and $T^2_{\pi(i)}$ can be selected 
independently for all pairs in Definition~\ref{def:pseudo-dist}.
Therefore, 
$Pdist(S_X,S_Y) \leq Dist(S_X,S_Y)$ holds.

Next we show that computation of
$Pdist(S_X,S_Y)$ can be reduced to the minimum weight perfect matching for
a bipartite graph.
From given $S_X$ and $S_Y$,
we construct a weighted bipartite graph $G(U,V;E)$ such that
$U  = \{u_1,\ldots,u_{m_2}\}$, 
$V  = \{v_1,\ldots,v_{m_2}\}$,
$E = U \times V$,
and 
\begin{eqnarray*}
w(u_i,v_j) = \left\{
\begin{array}{ll}
dist(T^1_i,T^2_j) & \mbox{if $i \leq m_1$,}\\
dist_0(\epsilon,T^2_j) & \mbox{otherwise.}
\end{array}
\right.
\end{eqnarray*}
Then, it is straightforward to see that
$G(U,V;E)$ has a perfect matching and
the weight of the minimum weight perfect matching is equal to
$Pdist(S_X,S_Y)$.
Since it is well-known that the minimum weight perfect matching
can be computed in polynomial time (e.g., by Hungarian algorithm),
the proposition holds.
\QED

\section{Computational Experiments}

We conducted preliminary computational experiments to
examine the possibility of the practical applicability of the proposed approach.
We focused on the unordered tree cases because
many operators (e.p., `+'. `$\times$') satisfy the commutative law.
Accordingly, we computed $Dist(S_X,S_Y)$ and $Pdist(S_X.S_Y)$
using several systems of differential equations
and normal equations
obtained from 
the BioModels data repository (https://www.ebi.ac.uk/biomodels/).
Equations whose left hand side parts contained variables that did not appear
in the right hand side of any equation were ignored.

Here we notice that the algorithms given in Propositions \ref{prop:elementary-dist}
and \ref{prop:pseudo-dist} use
an $O(1.26^{n_1+n_2})$ time algorithm \cite{akutsu14} to
compute the edit distance between unordered trees without variables.
However, this algorithm is not practical.
Thus, we employed an integer linear programming (ILP)-based approach
proposed by Kondo et al. \cite{kondo14} in place of \cite{akutsu14},
where we used the unit cost model for the simplicity.
Since the same constant symbols may be used for different meanings in different
systems, 
all constants are treated as different ones in two input models.
For the ease of implementation,
we also employed a simple ILP-based method to compute
the minimum weight perfect matching.
All computations were done on a Linux server with Xeon(R) CPU E5-2620 CPU 
$\times$ 2 and 132GB memory, using
IBM(R) ILOG(R) CPLEX(R) Interactive Optimizer 12.7.1.0 as an ILP solver.

The results are summarized in Table~\ref{tbl:experiment}.
The number inside parentheses denotes the number of variables,
and "Time Over" means that computation did not finish within 20 minutes.
CPU time denotes the total user CPU time, where CPLEX is parallelized to
efficiently work on multi-core CPUs.
It is seen that $Pdist(S_X,S_Y)$ could be computed in reasonable CPU time
even for systems with more than 10 variables, whereas
$Dist(S_X,S_Y)$ could be computed only for systems with 
3 $\sim$ 5
variables.
It is reasonable from the high exponential factor given in 
Proposition \ref{prop:elementary-dist}.
It is to be noted that 
$Pdist(S_X,S_Y)$ could not be computed within the time limit
for the pair of BIOMD0000000330 and BIOMD0000000331 although
each system contains only 5 variables.
It is reasonable because large equations are included in
both models.
It is also seen that $Pdist(S_X,S_Y) \leq Dist(S_X,S_Y)$ holds
for all pairs $(S_X,S_Y)$ for which $Dist(S_X,S_Y)$ could be computed
within the time limit.
This result is reasonable from Prop.~\ref{prop:pseudo-dist}.
The results of these preliminary computational experiments
suggest that the ILP-based method for computing $Pdist(S_X,S_Y)$
may be useful to comparing moderate size biological systems not
containing large-scale differential equations.

\begin{table}
\caption{Results on Computational Experiments}
\label{tbl:experiment}
\begin{center}
\begin{small}
\begin{tabular}{|cc|ll|ll|}
\hline
 &  & \multicolumn{2}{c|}{$Pdist(S_X,S_Y)$} & \multicolumn{2}{c|}{$Dist(S_X,S_Y)$} \\
$S_X$ & $S_Y$ & CPU time & Dist & CPU time & Dist \\
\hline
BIOMD0000000274 (3) &
BIOMD0000000680 (4) &
36,6 sec. & 50.0 & 55.9 sec & 55.0 \\

BIOMD0000000679 (4) &
BIOMD0000000252 (4) &
46.0 sec. & 40.0 & 43.7 sec & 46.0 \\

BIOMD0000000274 (3) &
BIOMD0000000330 (5) &
11 min. 29 sec. & 115.0 & 9 min. 8 sec. & 121.0 \\

BIOMD0000000274 (3) &
BIOMD0000000298 (11) &
8 min. 37 sec. & 246.0 & Time Over & N/A\\

BIOMD0000000252 (4) &
BIOMD0000000377 (14) &
17.2 sec. & 118.0 & Time Over & N/A\\

BIOMD0000000298 (11) &
BIOMD0000000377 (14) &
10 min. 48 sec & 253.0 & Time Over & N/A\\

BIOMD0000000330 (5) &
BIOMD0000000331 (5) &
Time Over & N/A & Time Over & N/A\\

\hline
\end{tabular}
\end{small}
\end{center}
\end{table}

\section{Concluding Remarks}

In this article, we proposed tree edit distance with variables,
motivated from needs for comparison of non-linear biological systems.
We analyzed the computational complexities of
several variants/subcases of the problem, and found that
the complexity substantially changes according
to introduction of variables.
We also extended the tree edit distance problem with variables
for computing the distance between two systems of first-order differential
equations.
Furthermore, we developed exponential-time algorithms for
the distance problems for unordered trees and systems.
However, the exponential factors are too high.
Therefore, much efficient exponential-time algorithms, 
parameterized algorithms, and/or approximation algorithms should be
developed.

From a practical viewpoint, some improvements should be made.
We employed the ILP-based method given in \cite{kondo14}
for computation of the unordered tree edit distance.
However, it is not necessarily the fastest one.
Indeed, the same research group developed an improve ILP-based method
\cite{hong17}.
Therefore, such an improved method should be considered.
In the computation of substitutions,
we employed a kind of exhaustive search.
However, this part could be improved by employing ILP or some other 
technique(s).
Therefore, the use of such techniques should also be considered.
Another important future work would be to assess the usefulness of
the proposed distances in measuring biological systems.
To this end, we need to conduct computational experiments using much more
data from the BioModels data repository, which again
requires developments of more practical methods.

\section*{Acknowledgment}

We are also grateful to the members of Sato lab at ATR and Karydo TherapeutiX,
Inc. for advice and discussion throughout the course of this work.
This work was supported in part by JSPS KAKENHI Grant Numbers
JP18H04113 (T.A.), JP17H06003 (N.N.), JP19H05422 (N.N.),
JST ERATO Grant Number JPMJER1303(T.N.S),
Nakatani Foundation(T.N.S) and AMED under Grant Number JP21he2102002 (T.N.S).

\end{document}